\documentclass[conference]{IEEEtran}
\IEEEoverridecommandlockouts
\usepackage{cite}
\usepackage{amsmath,amssymb,amsfonts}
\usepackage{algorithmic}
\usepackage{caption}
\usepackage[ruled,linesnumbered]{algorithm2e}
\usepackage{graphicx}
\usepackage{textcomp}
\usepackage{xcolor}

\def\BibTeX{{\rm B\kern-.05em{\sc i\kern-.025em b}\kern-.08em
    T\kern-.1667em\lower.7ex\hbox{E}\kern-.125emX}}
\begin{document}

\title{Hadath: From Social Media Mapping to Multi-Resolution Event-Enriched Maps
}

\author{\IEEEauthorblockN{Faizan Ur Rehman\textsuperscript{1,2}, Imad Afyouni\textsuperscript{3}, Ahmed Lbath\textsuperscript{1}, Saleh Basalamah\textsuperscript{4}}
	\IEEEauthorblockA{\textsuperscript{1}\textit{Laboratoire d'Informatique de Grenoble, University of Grenoble Alpes}, France \\
		\textsuperscript{2}\textit{Science and Technology Unit, Umm Al-Qura University}, Saudi Arabia \\
		\textsuperscript{3}\textit{Department of Computer Science, University of Sharjah}, UAE \\
		\textsuperscript{4}\textit{College of Computing and Information Systems, Umm Al-Qura University}, Saudi Arabia \\
		\{faizan-ur.Rehman, ahmed.Lbath\}@univ-grenoble-alpes.fr, iafyouni@sharjah.ac.ae, smbasalamah@uqu.edu.sa }

}

%\and
%\and
%\and
%\and
%\and

\maketitle

\begin{abstract}
Publicly available data is increasing rapidly, and will continue to grow with the advancement of technologies in sensors, smartphones and the Internet of Things. Data from multiple sources can improve coverage and provide more relevant knowledge about surrounding events and points of Interest. The strength of one source of data can compensate for the shortcomings of another source by providing supplementary information. Maps are also getting popular day-by-day and people are using it to achieve their daily task smoothly and efficiently. Starting from paper maps hundred years ago, multiple type of maps are available with point of interest, real-time traffic update or displaying micro-blogs from social media. In this paper, we introduce Hadath, a system that displays multi-resolution live events of interest from a variety of available data sources. The system has been designed to be able to handle multiple type of inputs by encapsulating incoming unstructured data into generic data packets. System extracts local events of interest from generic data packets and identify their spatio-temporal scope to display such events on a map, so that as a user changes the zoom level, only events of appropriate scope are displayed. This allows us to show live events in correspondence to the scale of view – when viewing at a city scale, we see events of higher significance, while zooming in to a neighbourhood, events of a more local interest are highlighted. The final output creates a unique and dynamic map browsing experience. Finally, to validate our proposed system, we conducted experiments on social media data.  

\end{abstract}


\section{Introduction}

Digital mapping applications to use live data to find directions, traffic congestion states or places of interest are used by users' to achieve their task in smart manners. These available maps have limited knowledge about Points of Interest (PoIs), real-time traffic update and weather information. While some of the map systems display micro-blog from social media such as Flickr images (https://www.flickr.com/map/) or real-time tweets\cite{MagdyEtAl2014}, there are still lot of social media data available which is increasing day-by-day. Analysing these data can provide deep insights about live surrounding events or any unusual happenings, given that a lot of relevant spatio-temporal information is embedded in social media streams. Social events of interest usually include gatherings, concerts, incidents, job announcements, or natural disasters, among others. Detecting or predicting such events in real-time can leverage new ways for exploring cities with dynamic content generated by the live communities in the surroundings, thus helping decision makers and authorities in providing context-aware intelligent services to their audience. Let's take an example of a small family visiting Paris city for the first time, and is interested in planning the best weekend trip to visit the top attractions including historic places with entry discounts, museums, musical concerts, good Japanese food with family promotions, and other kid- and family- friendly activities. An optimized trip to reach such attractions in a minimal time, mainly requires gathering live events and announcements for that weekend, optimized routes that take opening hours and transportation facility, as well as crawling and displaying relevant PoIs along with their reviews and ratings. This type of a query is difficult to answer on current mapping systems. A native approach would require an expert user who can look into traditional search engines to look for musical events within this period, and for Japanese restaurant reviews, and then try to manually best-match the different attractions while satisfying the strict time constraint.

\begin{figure}
	\centering
	\includegraphics[scale=0.32]{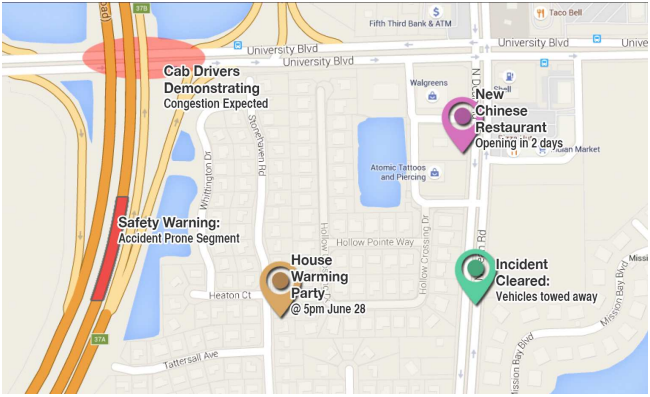}
	\caption[Caption for LOF]{Conceptual illustration of Event-Enriched Maps: Findings automatically discovered from live	streams, such as a restaurant opening, a neighborhood party, an accident prone road segment, warnings on demonstrations and emergency cases.}
	\label{fig:teaser}
\end{figure}

We believe a major additional functionality that can be integrated into maps will be displaying \emph{live and historical events}, extracted dynamically from user-generated content or crowd-sourced data. If intelligent mapping systems can discover relevant information from these unstructured data sources, the map browsing experience can be enriched significantly. For example, a spike in tweets talking about food at a particular location coupled with new Foursquare check-ins, can indicate the opening of a new restaurant (see Figure \ref{fig:teaser}). In contrast to traditional digital maps, \emph{smart mapping systems} can discover new content automatically by identifying new points of interests, events, or findings that were not specifically entered to the map.

Within this context, there is a real opportunity to enrich current maps with knowledge extraction tools that take advantage of information retrieval, data management, and sentiment analysis techniques. Analyzing crowdsourced data can provide deep insights about surrounding Events of Interest (EoI). For instance, with the explosive growth in size of microblog data (e.g., Twitter, Flickr, and Yelp), fruitful insights can be extracted and displayed (examples of discovered findings are illustrated in Figure 1). However, designing an efficient and scalable system that extracts live events and infers their spatial and temporal scopes, so that they can be displayed in a clear, non-cluttered manner, remains a challenging task.

The challenge in identifying new information to display on the map is multidimensional. First we need to infer map-worthy events and new places of interest from diverse live streams. This is a challenge in natural language understanding and context extraction. Secondly, to display such events on a map, their significance and spatio-temporal extent or scope must be established, so that as a user changes the zoom level, only events of appropriate scope are displayed. For example, a soccer match may be displayed at the city scale, the opening of a new restaurant at sub-urban scale and a house-warming party at the neighborhood scale. Thus, not only is it necessary to extract the events themselves, but also to establish their significance and how they span with respect to space and time, so that they can be displayed appropriately in a clutter-free manner. Finally, all of this has to be done in real-time so that live streams can be handled.

To address these challenges, we present a major extension of the work \cite{faizanEDBT2017,RehmanEtAl2017,RehmanEtAl2017a} by introducing a fully-fledged system named \textit{Hadath} that handles multiple types of data sources, and develops algorithms for an efficient extraction, clustering, and mapping of live crowd-sourced events. Our approach aims at providing an efficient and scalable framework for the management of a large number of microblogs that are disseminated worldwide, by employing a multidimensional in-memory indexing scheme, and a hierarchical clustering technique of candidate data points. Hadath digests incoming streams into a unique data packet format; and uses an approximate string matching technique (i.e., Tf-Idf and Cosine similarity) to extract candidate packets with an event corpus in order to identify potential event classes and properties. Moreover, the system develops an unspecified topic detection method that extracts spatio-temporal peaks and unusual happenings based on the occurrence score and diffusion sensitivity of topics. Local events that are extracted within limited spatial ranges, as calculated by the spatial index, can then be aggregated with similar events in neighboring areas in a hierarchical manner, so that the spatial and temporal scope of that aggregated event can be determined. Clustering of events is performed depending on the spatial and temporal dimensions, as well as the cosine similarity between related packets. Inferring the scope for a given event helps determining the map zoom level(s) for which this event should be displayed; thus providing an effective and smooth browsing experience of dynamic events.

The remainder of this paper is as follows. Section II presents the related work from several perspectives. Section III highlights the system overview of \textit{Hadath}, and then details the main architecture components. Section IV presents the implementation details and results; while Section V draws conclusions and discusses future challenges.

\section{Related Work}
Enriching maps with high-level extracted knowledge in real-time is of key interest to many areas of research, including real-time recommendation systems. Leveraging publicly available data allows for extracting up-to-date information about surroundings, thus enriching conventional spatio-temporal queries. This section highlights the state-of-the-art on topics related to existing mapping technologies and performance and scalability.

\subsection{State-of-the-art mapping technologies}

The use of digital maps is tremendously increasing with the aim of sharing preeminent information about current locations and spatial characteristics of the surroundings. Researchers, authorities, and industries generate thousands of map-based analytics every year to meet their social and economic needs\cite{krygier2011making}. Moreover, map giants including Google, Yahoo, Tomtom and Bing provide dynamic layers of traffic updates such as jams, accidents, and congestions to help users in their navigation needs. %Google Maps itself is used by 54\% of smartphone users which makes it among the most popular applications \footnote{http://mashable.com/2013/09/27/google-statistics/}.

\begin{figure*}
	\centering
	\includegraphics[scale=0.45]{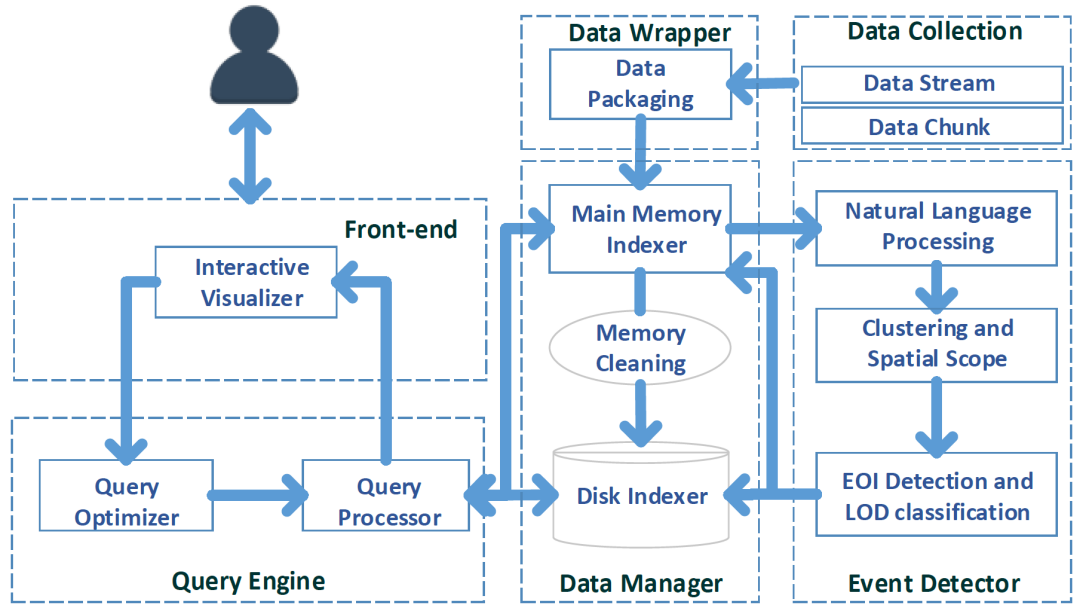}
	\caption[Caption for LOF]{Hadath Architecture}
	\label{highlevel}
\end{figure*}

Today's maps are often crowd-sourced, and make use of \textit{`Volunteered Geographic Information (VGI)'} \cite{Goodchild2007}, where users can seed maps with their own information. In addition, `Live Maps' now contain data that is updated in real-time. For example, live updates of bus schedules, traffic conditions, restaurant opening hours, and road accidents can be displayed on \textit{Google Maps}, and \textit{Waze}, among others. With the wide spread of social networks, people started to post their own social contributions on live maps, such as Foursquare check-ins\footnote{http://www.4sqmap.com/checkins/map}, Flickr images\footnote{https://www.flickr.com/map}, and tweets (Taghreed \cite{MagdyEtAl2014}, MapD\footnote{https://www.mapd.com/demos/tweetmap/}), Points of Interest (PoIs) reviews \footnote{https://www.yelp.com/wordmap/}, and news RSS (Rich documentation framework Site Summary) feed \cite{Teitler2008}. Moreover, Natural Language Processing (NLP) techniques were embedded to extract spatially-referenced emergency scenarios\cite{PMID:29186080} and news from online newspapers and tweets \cite{SametEtAl2014}.

Our proposed \textit{Hadath} system will enhance existing map solutions by detecting and displaying events of interest such as incidents, disasters, concerts, elections and parties. The novel \textit{Hadath} system will help in decision making and can be used by market firms, city development authorities, trip planners, and traffic departments, to interactively visualize past, current, and (near) future events on the map, along with their spatio-temporal.

\subsection{Performance and Scalability Perspectives} 

With the large volume of incoming streams, data indexing and the distributed processing of data represent an essential part of any system that implements \textit{`enriched maps'}. A significant challenge here consists in managing input data streams, cleaning and as well as extracted events for efficient processing and retrieval. Browsing \textit{Hadath} requires a smooth and fast panning and zooming capabilities. Events of different levels of abstraction are shown on the fly depending on the user navigation behaviour. Both real-time and historical data (up to a certain threshold) need to be indexed and processed for extracting the different categories of events. Consequently, such a system should provide support for both main-memory and disk-resident indexes.

Several works have presented systems that visualize geo-tagged social streams on maps, such as Flickr images\footnote{https://www.flickr.com/map}, tweets \cite{MagdyEtAl2014}, Yelp reviews, and spatially-referenced news \cite{SametEtAl2014, Sankaranarayanan2009}. Taghreed system \cite{MagdyEtAl2014} provides a mechanism to querying and visualizing tweets on maps by using spatio-temporal indexing techniques to run in real-time on current and historical data. Other works have been presented to detect communities of interest, event popularity, recommend optimized paths based on traffic constraints or to forecast upcoming events \cite{Gutierrez2015, Papadopoulos2012, RehmanEtAl2015}. The authors in \cite{TEDAS_ED2012} developed a system to detect crime and disaster events from tweets along with their spatial and temporal patterns. NewsStand \cite{SametEtAl2014} is a scalable system that extracts news from RSS feeds and visualize them on a worldwide map. Furthermore, the system can apply spatio-temporal and keyword-based filtering of news. However, this system displays news by only ranking them based on the number of views, without clustering events of interest based on their spatio-temporal extents. TwitterStand \cite{Sankaranarayanan2009} extends NewsStand to identify tweets related to late-breaking news and visualizes those tweets on maps. Moreover, the authors used a naive Bayes classifier to remove noise, i.e. tweets that are not related to news, and a leader-follower clustering algorithm to cluster tweets that belong to the same news. Consequently, although TwitterStand appear to be the closest work to our proposed system, it lacks a thorough understanding of any kind of unusual happenings as it only focuses on news. Also, this system does not consider the multi-scale nature of detected events.

\section{System Design}
In Hadath system, the context of what to display is set by the spatial extent of the detected events. When viewing the entire city, events that have a global interest should be displayed. As the user zooms in, events of progressively narrower scope must be displayed. For example, a soccer match can be of interest for the global scale, whereas a wedding may be of interest only at the district scale. However, even in the case of a wedding, the same event may need to be displayed on higher levels of abstraction if it involves lots of streaming input from other neighborhoods or cities, as may be the case of a celebrity wedding. In addition, unlike existing systems where knowledge is extracted based on users' requests, the key principle behind \textit{Event-Enriched Maps} is to extract knowledge on the fly by digesting social data streams and to infer its spatial scope, whether it covers a neighborhood, town, county, state, national or international level. This section presents \textit{Hadath}: a novel map-based platform that collects social data streams from multiple sources, processes data to find Events of Interest (EoI) and visualizes detected events in correspondence to the scale of the view. Figure \ref{highlevel} illustrates an overview of our \textit{Hadath} architecture with the salient components, which are highlighted as follows.

\noindent- \textit{Data collection} module involves gathering data from multiple sources with different unstructured forms. This includes digesting data streams (e.g., Twitter, Instagram, Yelp) and data chunks (e.g., open government, news, historical tweets). Digesting data streams is performed by running crawlers that collects bulks of streams based on windows of a specified temporal extent. 
	\begin{figure}[htp!]
	\centering
	\includegraphics[scale=0.31]{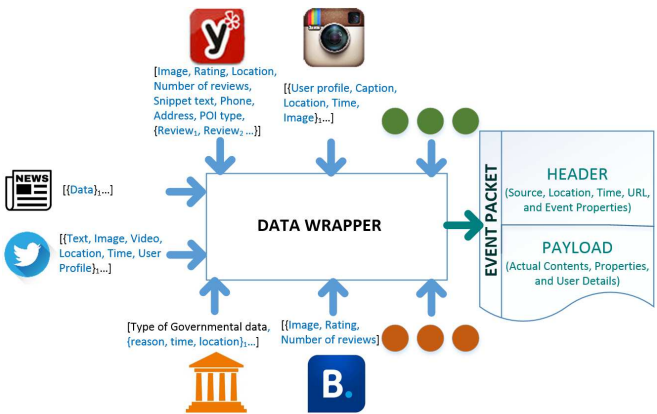}
	\caption[Caption for LOF]{Wrapping multiple sources of data into a generic composite event packet}
	\label{wrappper1}
\end{figure}

\begin{figure}[htp!]
	\centering
	\includegraphics[scale=0.3]{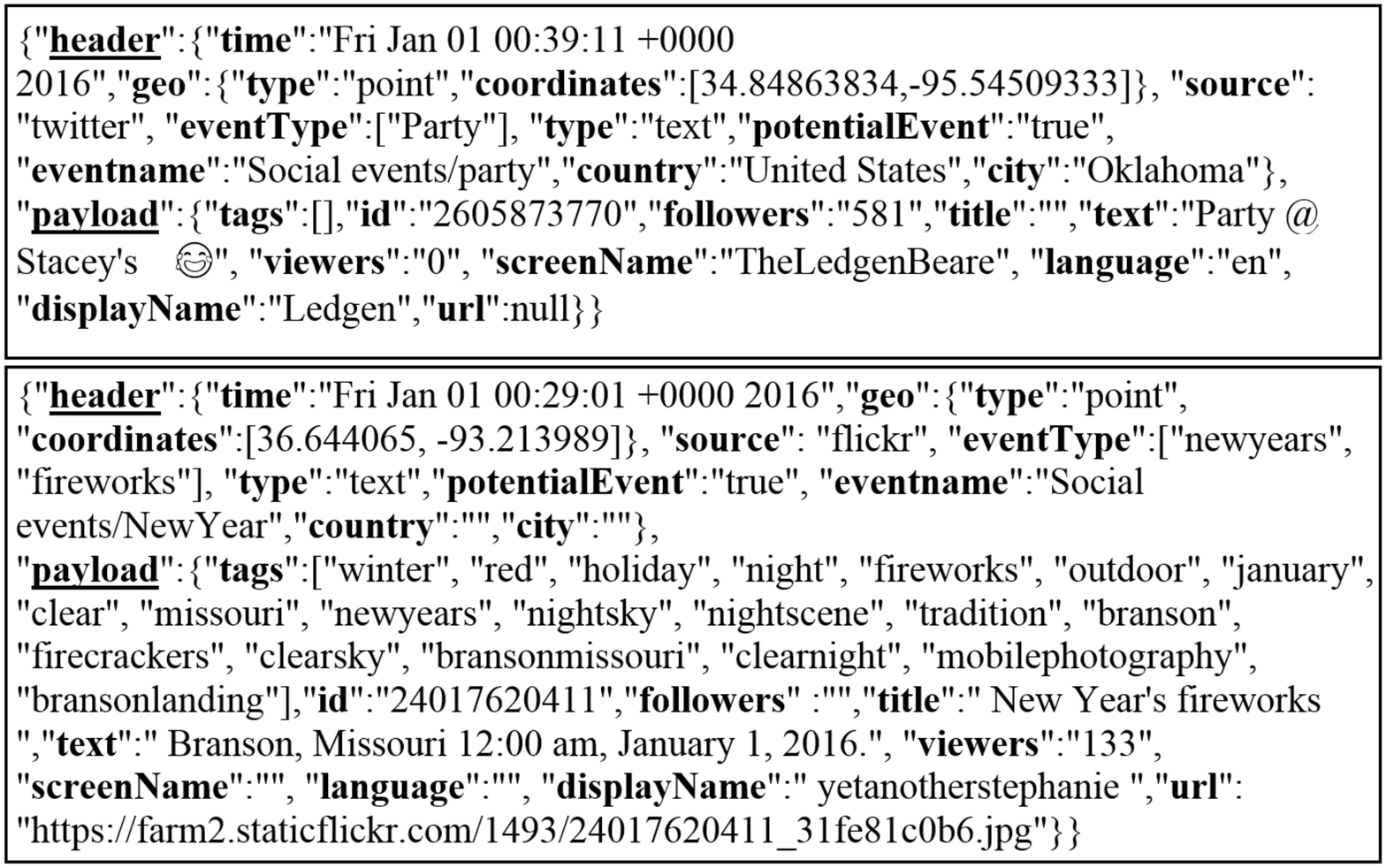}
	\caption[Caption for LOF]{Sample Data Packet from Twitter and Flickr Streams}
	\label{datapacket}
\end{figure}
\noindent- The \textit{data cleaner and wrapper} provides an efficient and generic mechanism with the aim of allowing new data sources (e.g., Instagram) to be easily plugged, by supporting new crawlers at the data collection level without affecting the other processing components. Figure \ref{wrappper1} shows a conceptual example of data packets generated from multitude of data. Major tasks for the data cleaner and wrapper are: 1) to clean irrelevant fields and digest incoming streams into a unique data packet format. Like the analogy of TCP protocol, each data packet has a meta-data header, containing source, location, time, type, and a payload, containing the actual contents including user profile details. This allows our system to digest different types of data input, and to generate structured data from unstructured streams; 2) to use specified string matching technique that detect and match candidate packets with our event classifier corpus in order to identify potential event classes and properties. This approach helps us to extract relevant packets related to known event classes including social, disaster, religious, weather, job, traffic, sports, political/government and musical events ; and 3) to apply unspecified topic detection method that extract spatio-temporal peaks and unusual happenings based on the top frequent words. This approach helps us to detect unknown events that are not a part of corpus but their value are more than threshold at given time. To add new source in our \textit{Hadath}, we just need to add small piece of code with out impacting other components of the system. The header and payload are populated in different ways for different data sources. For example, `event properties' may be populated by the parsing hashtags, noun, verb in case of Twitter, and hashtags in case of a Flickr. Hence, a different data filter is written for each new source which is added to the system. This allows our system to digest different types of data input, and to generate structured data from unstructured streams. Figure \ref{datapacket} shows sample packets from Twitter and Flickr Stream. These packets are then processed to extract Events of Interest, and can be dealt with in a consistent manner by the higher layers of the system.

\begin{figure}[tp!]
	\centering
	\includegraphics[scale=0.22]{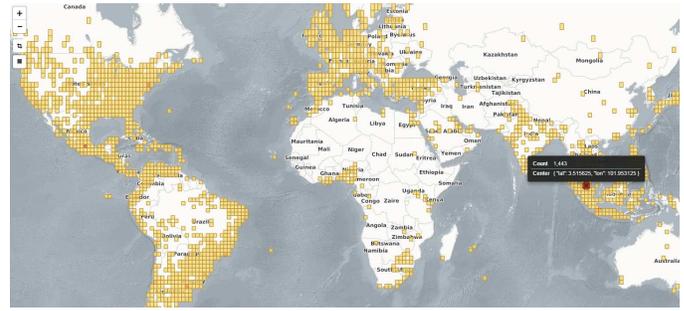}
	\caption[Caption for LOF]{A snapshot of indexed data packets at a fine level of the hierarchical tree}
	\label{quadtree}
\end{figure}

\begin{figure}[htp!]
	\centering
	\includegraphics[scale=0.38]{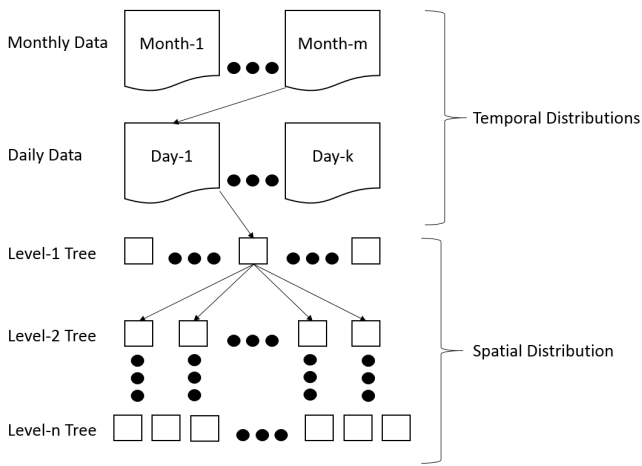}
	\caption[Caption for LOF]{Multi-level spatio-temporal index}
	\label{diskIndex}
\end{figure}

	\begin{figure*}[htp!]
	\centering
	\includegraphics[scale=0.45]{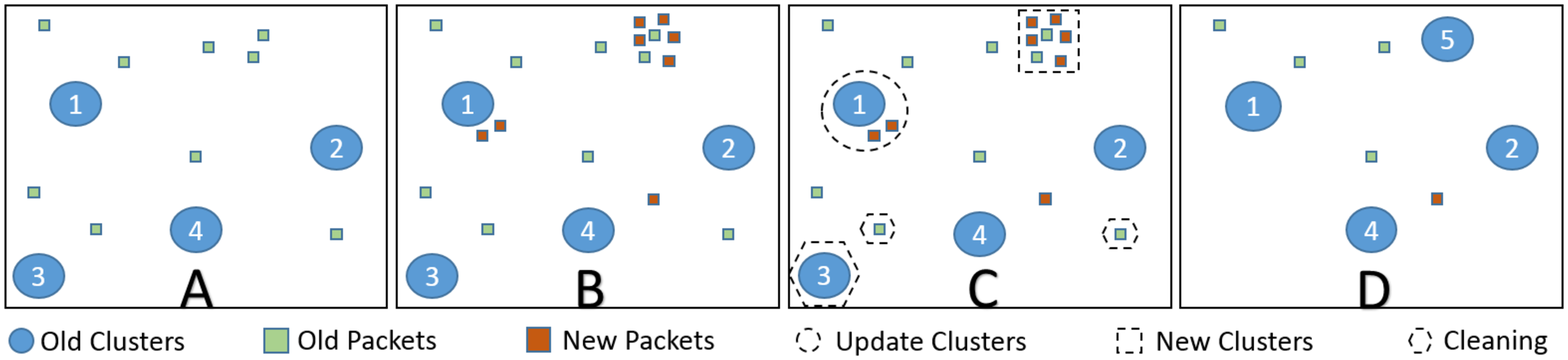}
	\caption[Caption for LOF]{Local EoI Detection within the leaf cells A) List of existing clusters and packets, B) shows arrival of new indexed data packets, C) Merging new packets within an existing cluster, forming a new cluster, and cleaning old packets and clusters based on a threshold value, D) resultant leaf cells }
	\label{eois1}
\end{figure*}

\noindent- The \textit{data manager} implements an in-memory spatial indexing scheme to allow an efficient and scalable access to data packets. The spatial index is a multi-resolution data structure (similar to a partial quad tree originally introduced in \cite{Samet1984}). Leaves in this data structure correspond to cells that represent the minimum bounding rectangles comprising data packets. Figure \ref{quadtree} displays a snapshot of indexed data packets at a fine level of the hierarchical tree, and with a single day specified as a time threshold. Cells are colored lighter to darker based on data packet counts; darker-colored cells are further expanded at deeper levels in the tree as compared to lighter-colored cells. \textit{Hadath} employs a big data mechanism that continuously process data packets within the different cells on several execution nodes. The manager also indexes detected EoIs in order to fetch them efficiently based on the map zoom level and scope. Using this multi-resolution indexing scheme, hierarchical clustering of events can be applied for efficient determination of their content and spatial scope. 

To get better understanding about the place, users' may also be interested in old EoIs apart from current or upcoming EoIs such as if a user wants to book an hotel in new city then she can browse nearby area and see what happened around in the past. To support EoIs for long periods, our system stores data from main-memory to disk based on main memory threshold or based on temporal threshold. However, the disk index is a bit different from the main memory with respect to temporal parameters. Figure \ref{diskIndex} shows an overview of the hierarchical disk index implementation in Hadath. To access EoIs efficiently, we distribute temporal data packets to monthly and daily bases. The monthly distribution stores month with year whereas daily distribution inside monthly distribution stores actual EoIs of particular day in a same pyramid manner explained in main memory section. For example, if a user requests data of june 2017, then the query processor needs to access 30 indexes inside the june month to fetch EoIs and keyword indexes.

The main task of \textit{memory cleaning} manager is to clean old data packets and EoIs from memory. Old data packets are removed on the basis of temporal aspects. We have used periodic approach to clean EoIs periodically in conjunction with the piggybacking approach over the querying process. Whenever an EoI happen, you can check all EoIs for that particular type and update it.
	
\noindent- After cleaning, wrapping and indexing of data packets, the \textit{event of interest detection} module starts from the base level within the multi-resolution data structure to detect events at a local spatio-temporal scope. The base level contains cells of a fine resolution, that are considered as leaves within the hierarchical pyramid data structure. Within each leaf cell, Hadath adopts the graph analogy where data packets are considered as nodes and the value of the `text similarity (TF-IDF)' between data packets is computed as the weight of the bidirectional edges. Data packets with a high text similarity value within each cell are clustered together using the graph-specific Louvain clustering algorithm. The Louvain algorithm \cite{Blondel08fastunfolding} is suitable in our approach as, unlike most of the other clustering methods, it does not require a prior knowledge of the minimum number of clusters. For unspecified events that are not matching our training corpus, this module detects frequent tags and keywords, in order to identify spatio-temporal peaks.

To create new event clusters within the leaf cells, an algorithm is developed so that similar packets within the local spatio-temporal scale are grouped together, or can be grouped within existing event clusters in the corresponding cell. Figure \ref{eois1} shows stages of an event of interest detector in the following four different stages. A) shows leaf level cluster with existing clusters (blue circle) and non clustered data packets (light green square) before the arrival of the newly indexed data packets. B) shows new indexed data packets (red square) in the same leaf level. Overall, it shows a list of old clusters, old packets and new Packets; C) New clusters are formed and old ones are updated with new packets. Clusters and packets are considered as nodes in the cell, and distances between them are computed based on the cosine similarity value. As a result, some packets are merged with existing clusters (dotted circle), and the combination of some old and new packets forms a new cluster (dotted square). Also, cleaning of old packets and clusters is performed at this level, based on their temporal threshold and TTL parameter (dotted hexagon). D) shows the final output of local event detection in leaf cells after creating new clusters, updating old clusters and cleaning of old packets and clusters.  
	
\begin{figure}[htp!]
	\centering
	\includegraphics[scale=0.38]{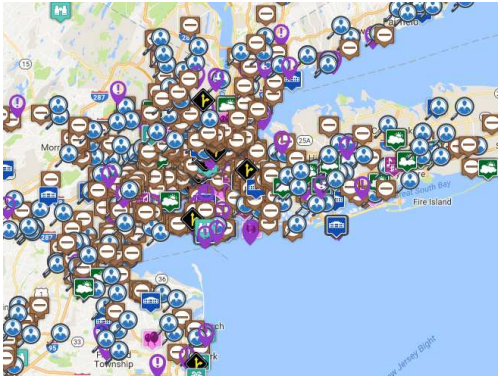}
	\caption[Caption for LOF]{Visualization of Event of Interest without spatial scope }
	\label{fig:viswithoutscope}
\end{figure}

\noindent- \textit{Clustering and Spatial Scope}. Visualization of events with diverse significance or impact should take into consideration the map spatial resolution, thus allowing a clutter-free and natural map browsing experience. Visualization of events with the same spatial resolution on maps does not make sense, since these events have different extents from spatio-temporal perspectives. Figure \ref{fig:viswithoutscope} illustrates diverse types of events as a result our event detection module. This forms an event portal that can be very useful for a city explorer or any decision maker, but due to clustered and overlapped events, it is difficult for someone to browse such types of maps without enhanced visualization of those events. For instance, events of someone's birthday cannot be displayed at a national level, except if that person is a celebrity, and that event was spread throughout the country.

As events can be discovered more efficiently on small-scale regions, a bottom-up approach for clustering close-by and similar events is developed, so that redundant events on different spatial resolutions can be aggregated, and their spatial scope can be upgraded. In order to update the granularity level and identify the spatial scope for a given event, \textit{Hadath} implements a hierarchical clustering technique that detects an event spatial extent starting from leaf nodes. This technique consists in aggregating nearby and matching events, so that redundant events in the sibling cells are merged, and a new cluster with a larger spatial scope is determined. This phase is repeated successively at higher levels of abstraction to incrementally increase events' spatial scope \cite{RehmanEtAl2017}.

\noindent- \textit{Query Engine }that has two components; a) query optimizer and b) query processor. The query optimizer creates best query plan based on map zoom level, spatial and temporal characteristics. The query processor executes the query plan in order to retrieve EOIs efficiently. \textit{Hadath}'s query engine supports efficient retrieval of raw streams and of pre-processed events based on the main querying attributes, that are, the spatial, temporal, and map level of detail. 
	
\noindent- The \textit{Front-end} provides a new dimension to existing maps by illustrating extracted knowledge from live streams in the form of live events with different spatial scopes and at different levels of abstraction. This allows us to show live events in correspondence to the map level of detail (LOD), that is, when viewing at a city scale, events of higher significance are displayed; whereas, when zooming in to a given neighborhood, events of a more local interest are highlighted. The final output creates a unique and dynamic map browsing experience.

\section{Implementation and Results}  

To validate our approach, Hadath was developed as a proof-of-concept system based on a big data framework towards efficient data management and visualization of events of interest on maps. Our methods were tested with more than 30 million geotagged tweets. The front-end is a map-based application that visualizes spatio-temporal events at different scales. The implementation is done in Java 1.7. We are using i7-4712 HQ-CPU @ 2.30GHz with 16GB DDR-2 RAM at the back-end for processing using the following libraries and software \cite{faizanEDBT2017}, 1) Ark-tweet-NLP \cite{Gimpel2011} for natural language processing on Twitter data, 2) Louvain clustering algorithm  \cite{Blondel08fastunfolding} to cluster the data with in the cell based on text similarity, 3) Taghreed crawler \cite{MagdyEtAl2014} to collect tweets, 4) Geo-Spatial Data aggregation \cite{AttaEtAl2016} for indexing the packets using grid-based approach in which we divide the world map into rectangular cells, and then grouping data points within each cell hierarchically using geohash. Geohash has a character value that helps in postulating the accuracy of the hash value and determining the location. For example, the latitude and longitude coordinates of 40.78, -73.96 fall within the geohash box of `dr' that it is a part of New York city, USA. Adding a character to the string `dr' will lead to more specified geographical subsets of the original string \cite{jiajun2014}. One of the advantages of the geohash technique is that it translates two-dimensional spatial queries into one-dimensional string search. Therefore, it can solve search queries with $O$(1) time complexity. The length of the geohash string is considered as the precision level for a specified zone.  As the geohash strings are shortened, less precise zones are covered, and 5) Elasticsearch for Apache Hadoop (ES-Hadoop 5.3) is a special library that allows for linking Hadoop jobs into Elasticsearch \footnote{https://www.elastic.co/products/hadoop}. ES-Hadoop serves as a link between Hadoop's big data analytics and Elasticsearch. Data nodes (for data storage) and TaskTracker (for data processing) are the two nodes that were installed. The following machine specification were adopted for Hadoop deployment. 2x1TB hard drives and 2 quad-core CPUs, running at 2.5GHz and 32GB of RAM. For in-memory operations and to leverage efficient storage and retrieval of data streams, Rabbit MQ server was installed as a RAM node.

\begin{figure}[htp!]
	\centering
	\includegraphics[scale=0.35]{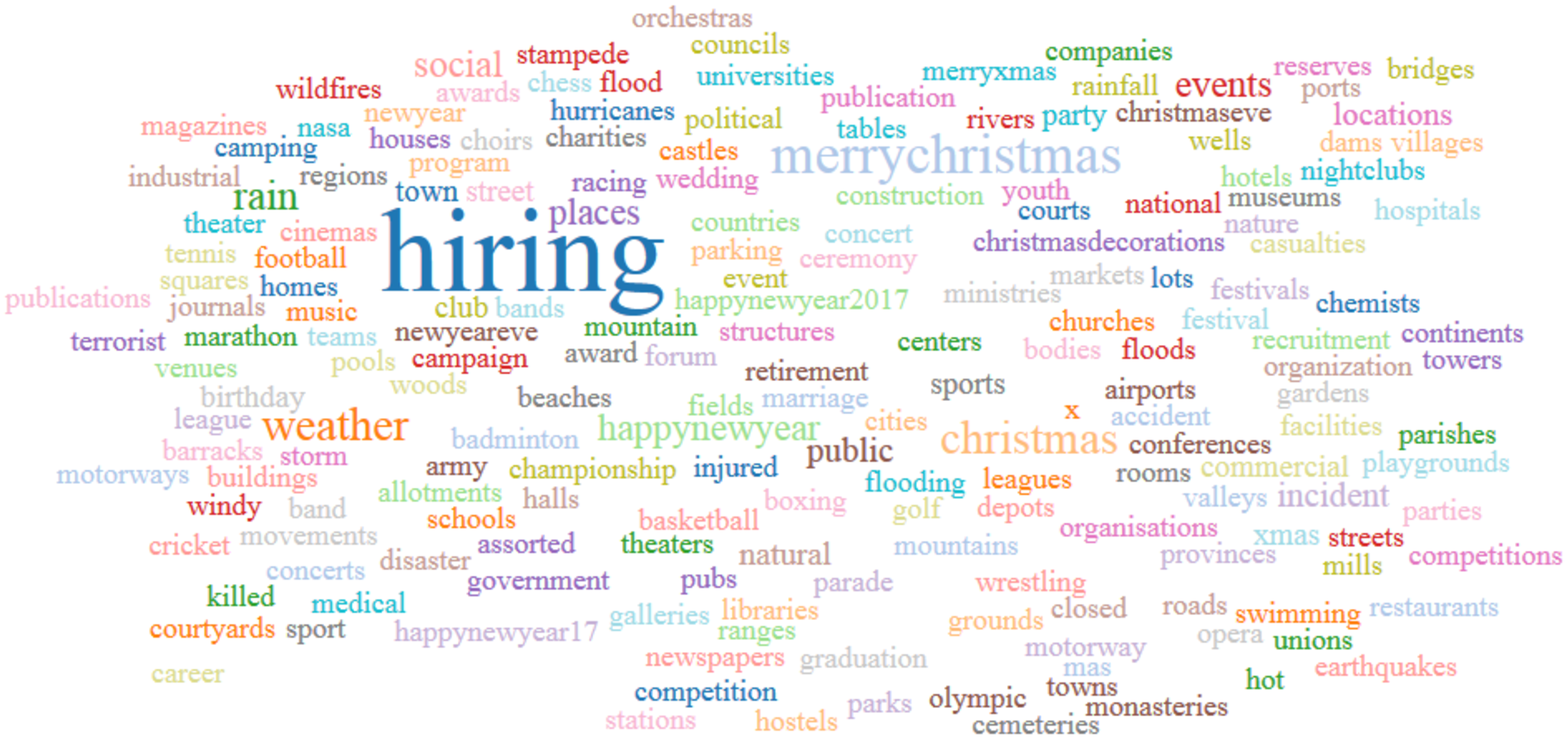}
	\caption[Caption for LOF]{Interactive Tag/Word Cloud of EoIs}
	\label{EoICloud}
\end{figure}

\subsection*{\textbf{Results}}
\begin{figure}[htp!]
	\centering
	\includegraphics[scale=0.3]{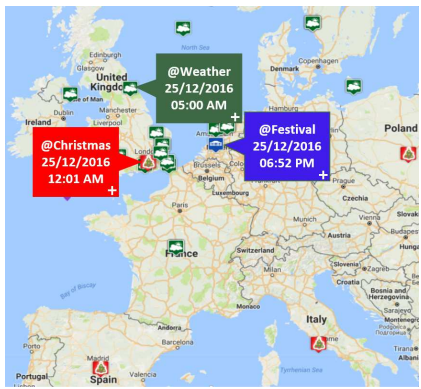}
	\caption[Caption for LOF]{High Absatrction Map View with Country Level Social Events}
	\label{Level01}
\end{figure}

\begin{figure}[htp!]
	\centering
	\includegraphics[scale=0.25]{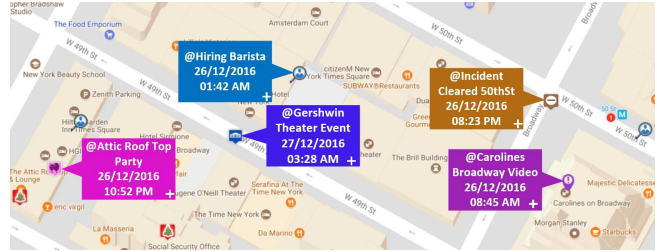}
	\caption[Caption for LOF]{Detailed Map View for the City of New York}
	\label{Level23}
\end{figure}

As previously mentioned, the aim of Hadath system is to users to interact with event-enriched maps smoothly, and browse events at different levels of details on a world wide map with an efficient panning and zooming capabilities. The following results intend to demonstrate the usage and performance of our prototype in discovering the spatio-temporal scope of social events. For instance, Figure \ref{EoICloud} illustrates an interactive tag/word clouds of events of interest that are dynamically adapted when changing the specified spatio-temporal scope (i.e., by zooming, panning or applying a rectangular range selection). The multi-resolution event-enriched map visualization, where events of global significance are displayed at higher abstraction levels and local significance are shown at lower abstraction levels are are shown in Figures \ref{Level01} and \ref{Level23}.

Figure \ref{Level23} displays a detailed map view at the neighborhood level in New York city. The nature of events is clearly distinguished at such a local scale, as we can find clusters of events discussing incidents, parties, hiring opportunities, or some other gathering in the district. Of course, such local event can be expanding further at higher levels of abstraction depending on how people interact with such disruptive news (e.g., a hiring announcement can, for instance, start at a local scale but a more global interest depending on the hiring company). It is also worth noting that some unspecified event, such as `Carolines Broadway Video', were discovered by utilizing our scoring technique to compute the keyword peaks while users are posting tweets on some unknown unusual happening. 

\begin{figure}[htp!]
	\centering
	\includegraphics[scale=0.4]{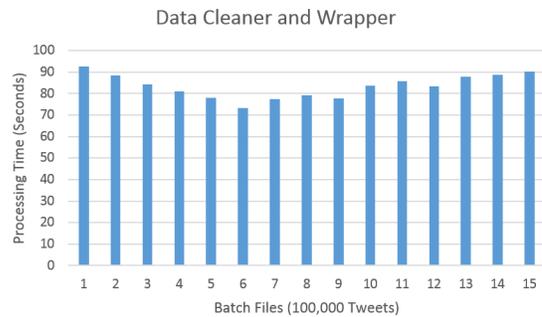}
	\caption[Caption for LOF]{Processing time for generating data packets from 100,000 tweets}
	\label{packetsTime}
\end{figure}

As illustrated in Figure \ref{packetsTime}, the processing time for generating indexed data packets after cleaning and wrapping of 15 batch files with each file containing 100,000 tweets. As demonstrated, a period of one minute to one minute and half is required for each bulk of data (i.e., every half an hour) to be indexed and merged with the existing clusters in the hierarchical spatial tree.

\begin{figure}[htp!]
	\centering
	\includegraphics[scale=0.32]{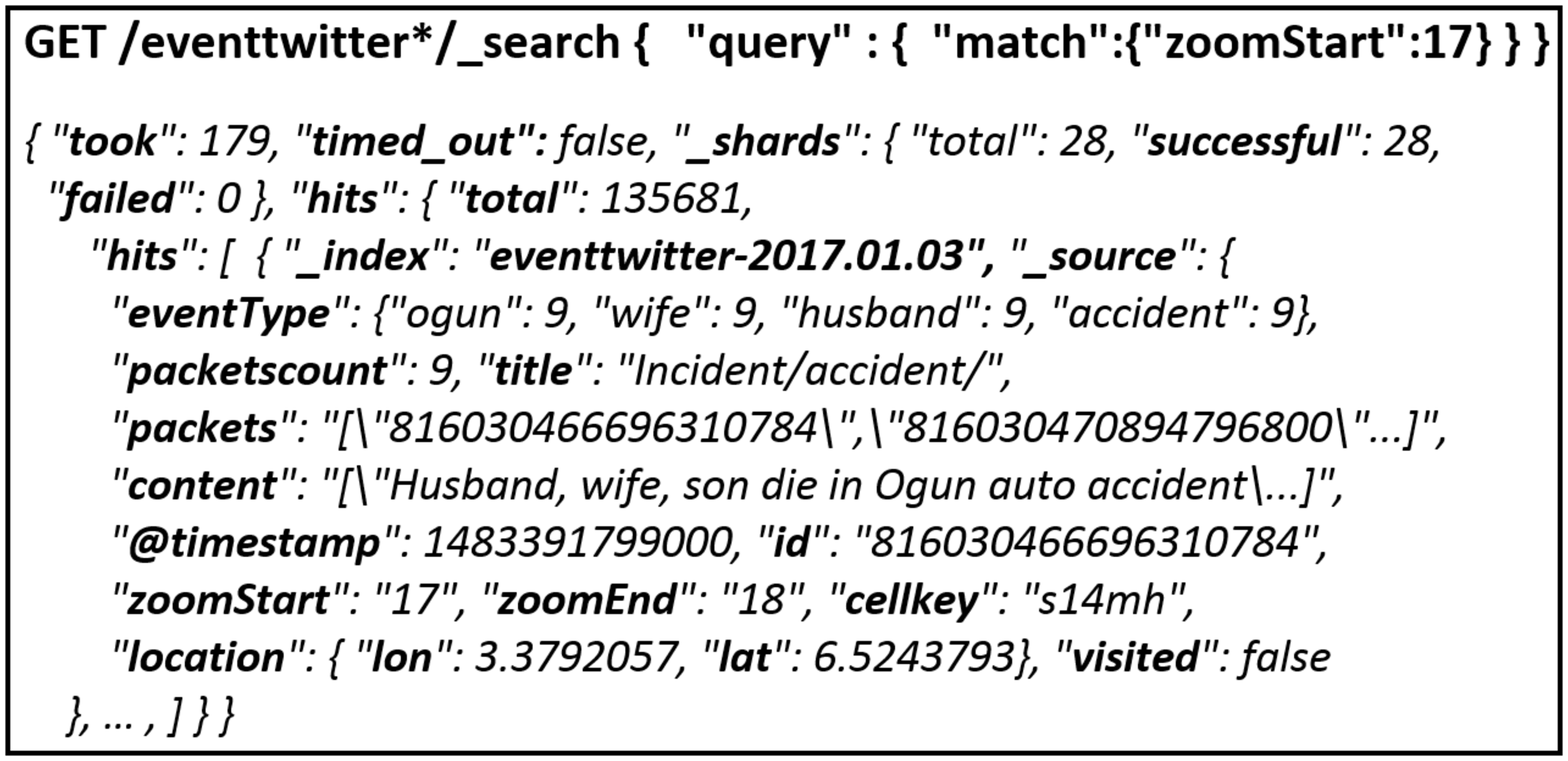}
	\caption[Caption for LOF]{Query and a sample EoI of zoom level 17}
	\label{IndexPacketKibana}
\end{figure}

Figure \ref{IndexPacketKibana} shows a query to fetch EoIs at zoom level $17$ and a sample result to demonstrate the structure of the EoI cluster. In the query result, we can see that it took $179$ milliseconds to return $135,681$ EoIs using $28$ shards with $100\%$ success rate. Each EoI contains, name of index as `\_index', properties of events as `eventType', total number of packets and their identifiers involved in clustering as `packetcount' and `packets' respectively, time stamp of EoI as `@timestamp', identifier of clustered EoI as `id', zoom level details of map as `zoomStart' and `zoomEnd', geohashing index of cell as `cellkey', centroid location of a cluster calculated by taking into consideration the location of all packets in clusters as `location' and a flag that is used to calculate vertical spatial scope as `visited'.

Its difficult to manually verify the accuracy of our system for the whole dataset on world wide map due to the huge amount of data. Therefore, to evaluate the correctness, we considered a geographical zone of size 156x156km which covers the city of Lagos, Nigeria's largest city. Lagos is one of the fastest growing city of the world and the most populous city in Africa, and also presented the highest number of packets for the evaluated time period. One full day of twitter data was considered for this experiment, with around 4 million raw data streams collected. We calculated the `Precision', `Recall', `Accuracy' and `F1' for each of the derived map scales as shown in Table \ref{tabPrecision}.

\begin{table}
	{\small\centering
		\begin{tabular}{||c | c | c | c | c | c | c| c | c||}  
			\hline
			\textbf{Scale} & \textbf{Precision}& \textbf{Recall}& \textbf{Accuracy}& \textbf{F1}\\  
			\hline\hline
			Local & 0.82 & 0.80 & 0.91 & 0.81 \\
			Neighborhood & 0.93 & 0.92 & 0.92 & 0.92 \\
			Sub-Locality & 0.95 & 0.98 & 0.99 & 0.96\\	
			Locality & 0.98 & 0.98 & 0.99 & 0.99\\
			City & 1 & 0.99 & 0.99 & 0.99\\	 		
			\hline
		\end{tabular}
		\caption{Precision, Recall, Accuracy, and F1-Score at different map scales}
		\label{tabPrecision}
	}
\end{table}

\section{Conclusions}

This paper introduces \textit{Hadath}, a system that builds multi-resolution event-enriched maps by handling social data streams, and by developing different algorithms for the efficient extraction, clustering, and mapping of live events.  The system can provide valuable knowledge from crowd-sourced data to authorities, market firms, event organizers, and end-users to help in decision making. In future, we plan to merge more data sources (e.g., Instagram, online newspapers) to increase correctness and conciseness of detected events. Furthermore, an extensive performance evaluation of the different techniques need to be conducted with respect to closely-related systems. We also need to handle historical data by developing statistical learning tools towards a better understanding of urban data and user behavior. We believe \textit{Hadath} can help in building the next generation of maps platform by intelligently extracting relevant knowledge from crowd-sourced data in real-time.

\section*{Acknowledgment}
We kindly acknowledge useful suggestions from Prof. Mohamed F. Mokbel, University of Minnesota, USA and Dr. Sohaib Khan, Umm Al-Qura University, Saudi Arabia.  We would like to thank Shahbaz Atta for his development support in big data. %Also,  we would like to thank Science and Technology Unit (STU), and Wadi Makkah Technology Innovation Center at Umm Al-Qura University, Saudi Arabia for providing the necessary resources.

\bibliographystyle{IEEEtran}
\bibliography{biblio}

\end{document}